\documentclass[preprint]{aastex}
\def\Vec#1{\mbox{\boldmath $#1$}}
\begin{document}
\title{Dipole anisotropies of IRAS galaxies and \\
the contribution of a large-scale local void}
\author{Kenji Tomita}
\affil{Yukawa Institute for Theoretical Physics, Kyoto University,
        Kyoto 606-8502, Japan}
\email{tomita@yukawa.kyoto-u.ac.jp}
\begin{abstract}
Recent observations of dipole anisotropies show that the velocity of
the Local Group ($\Vec v_{\rm G}$) induced by the clustering of IRAS galaxies
has an amplitude and direction similar to those of the velocity of
Cosmic Microwave Background dipole anisotropy  ($\Vec v_{\rm CMB}$),
but the difference  $\vert \Vec v_{\rm G} - \Vec v_{\rm CMB} \vert$ is 
still $\sim 170$ km/s,
which is about $28 \%$ of $\vert\Vec  v_{\rm CMB} \vert$.  
Here we consider the possibility that the origin of this difference comes 
from a hypothetical large-scale local void, with which we can account for the
accelerating behavior of type Ia supernovae due to the spatial
inhomogeneity of the 
Hubble constant without dark energies and derive the constraint to the model 
parameters of
the local void. It is found as a result that the distance between 
the Local Group and the center of the void must be  $(10$ -- $20) h^{-1}$ 
Mpc, whose accurate value depends on the background model parameters.
\end{abstract}


\section{Introduction}
\label{sec:level1}

From the observed dipole component of temperature anisotropies in the 
cosmic microwave background radiation (CMB), our peculiar motion 
relative to the CMB frame has been determined because the dipole
anisotropy is connected with the motion 
by a Doppler effect. The velocity of the Local Group relative to the
CMB frame is found to be 
\begin{equation}
  \label{eq:a1}
{\Vec v_{\rm CMB}} = (-25.2, \ -545.4, \ 276.5) =\ 612.0\ (-0.041,\ -0.89,\ 
0.45)\ {\rm km \ s}^{-1},
\end{equation}
where the $x$-axis and $z$-axis are towards the Galactic center and
the Galactic pole, respectively, and the total velocity $\vert {\Vec
v_{\rm CMB}}\vert$ is  
$612 \pm 22$ in the direction of $(l, b) = (268 \pm 3, 27 \pm 3)$ \
\citep{rr00,law99}.

The peculiar motion is, on the other hand, caused gravitationally by the
inhomogeneous matter distribution around us. Unitil now the latter velocity of
the Local Group ${\Vec v_{\rm G}}$ has been derived on the basis of
linear gravitational instability using matter distributions that 
have been taken from
 optical survey of galaxies, the IRAS galactic redshift survey
and cluster surveys, for a long time \citep{yahil86,meiks86,lah87,
rr90,strauss92,schmol99,branch99,rr00}. 
The problem of the convergence of the velocity has also 
been discussed in many studies \citep{juszk90,scaram94,plion98}.

The most recent determination of ${\Vec v_{\rm G}}$ was given by
\citet{rr00} using the Point Source Catalogue (PSCz) IRAS galactic
redshift survey, 
 which was analyzed out to a distance $300 h^{-1}$ Mpc \ ($H_0 = 100
h$ km s$^{-1}$ Mpc$^{-1}$):
\begin{equation}
  \label{eq:b15}
 {\Vec v}_{\rm G} \equiv {\Vec v}({\Vec r}_{\rm LG}) =\ 697.7\ 
{\Vec n}_{\rm G} \ {\rm km\ s}^{-1},
\end{equation} 
where ${\Vec n}_{\rm G} \equiv (-0.257,\ -0.811,\ 0.482)$ \ 
represents a directional unit vector. 
The above two velocities ${\Vec v_{\rm
CMB}}$ and ${\Vec v_{\rm G}}$ have similar amplitudes and directions,
but, when inhomogeneities out to $300 h^{-1}$ Mpc are taken into
account, the difference $\vert \Vec v_{\rm G} - \Vec v_{\rm CMB} \vert$ 
reaches $\sim 170$ km/s, which is about $28 \%$ of $\vert\Vec v_{\rm
CMB} \vert$. The origin of this difference may come from a
hypothetical object behind the Galactic center, inhomogeneous biasing
or nonlinear correction to perturbation equations, as was indicated by
Rowan-Robinson et al. (2000).      

In this paper we investigate the contribution of a hypothetical local
void in which the inner Hubble constant ($H_0^{\rm I}$) is larger than 
the outer Hubble constant ($H_0^{\rm II}$) and the radius of the
boundary is $\sim 200 (h^{\rm I})^{-1}$ Mpc, where $H_0^{\rm I} = 100
 h^{\rm I}$ km s$^{-1}$ Mpc$^{-1}$ \ (see Fig. 1). It is assumed that
we are near the center of the inner region \ (e.g. \citet{tom00a,
tom00b}).
  
\bigskip
\centerline{//\ Fig.1 \ //}
\bigskip

Here let us briefly describe the motivations for introducing the
local void and the significance of its  model parameters.
Firstly, the observed values of the Hubble constant appear to be inhomogeneous:
the local value ($H_0^{\rm I}$) in the neighborhood is about 72 km
s$^{-1}$ Mpc$^{-1}$ \citep{sakai00, free01}, but the global values
($H_0^{\rm II}$) in remote regions are smaller than $H_0^{\rm I}$,
by factors  
which reach even $20 \%$. The global values have been derived
using gravitational lensing \citep{keet97,courb97,fass99,
will00,tada01, koch02a,koch02b, koch02c} and the 
Sunyaev-Zeldovich effect \citep{reese00}, although they depend on
cosmological models and include some uncertainties.
This inhomogeneity of the Hubble constant strongly suggests the
existence of a local void (on scales with the redshift $z < 0.1$), 
because, if $H_0^{\rm I} > H_0^{\rm II}$
locally, the central region must necessarily be underdense with
respect to the total matter density. A local void on scales $\sim 70$
Mpc has once be studied by \citet{zeh98} in terms of the Hubble
bubble. At present, the existence of an
underdense region on such scales is not clear from 
optical observations, athough it has so often been suggested on
scales $\sim 300$ Mpc
\citep{mari99, marz98, folk99, zucca97}, and so we should expect 
the situation that the distribution of luminous matter (such as
galaxies and clusters) is nearly homogeneous, while that of dark matter 
is locally inhomogeneous. This situation, which seems to be
contradictory may be realized through the
difference of the feedback system in hierarchical galaxy formation 
between the central (local) region and the outer (global) region
\citep{tom02},
since in the low-density region, more galaxies are produced owing to
less supernova explosions and photoionization, than in the
high-density region \citep{kauf99, ben01, som01}.  

Second, the accelerating behavior of high-redshift supernovae (SNIa) 
was found by two groups, the High-$z$ SN Search Team \citep{schmidt98,
riess98, riess99} and the Supernova Cosmology Project \citep{perl99}. 
Its origin has usually been explained by considering the role of nonzero
cosmological constant  
$\Lambda$ or dark energies in homogeneous cosmological models. Because 
the existence of nonzero $\Lambda$ or dark energies is not evident
theoretically, however, the search for other explanations may be
significant. In the model with $H_0^{\rm I} > H_0^{\rm II}$, on the other
hand, the
underdense region (the local void) plays the role of vacuum, and
supernovae there show accelerating behaviors, even though we have no
$\Lambda$ and dark energies. In our recent papers \citep{tom01a,
tom01b}, it was shown that, for larger $(H_0^{\rm I} - H_0^{\rm II})
/H_0^{\rm II}$, the magnitude ($m$) and redshift ($z$) relation of
high-redshift supernovae can be well reproduced in the models with smaller 
$\Lambda$, and, if $(H_0^{\rm I} - H_0^{\rm II})/H_0^{\rm II} = 0.2$, the 
relation can be quantitatively explained in the model which reduces to 
the Einstein-de Sitter model. The best radius of the boundary for 
fitting  
the $m-z$ relation is found to be $\sim 200$ Mpc. Moreover the latest
data of a supernova with $z \sim 1.7$ \citep{riess01} are naturally 
consistent with
this last model with $\Lambda = 0$ \citep{tom01b}. For cosmological
parameters in the inner region, we take the low-density value
($\Omega_0^{\rm I} \sim 0.3$), corresponding to the statistical
analyses for local observations of galaxies and clusters.
  In the outer region, spatially flat 
models are chosen in accord with the recent results of CMB
observations \citep{lang, stomp, pryk}, and our models with small or
vanishing $\Lambda$ are found to be compatible with their data in the
weak prior. 

Third, bulk flow on the scale more than 100 Mpc was found by 
by \citet{hud99} and \citet{will99}. It is impossible to account for
its appearance as one of the linear perturbations in homogeneous
cosmological models. In a model with a local void on scales $\sim
200$ Mpc, however, we can interpret the bulk flow as the nonlinear 
deviation of
the local flow from the global expansion of the universe. 
This point is also one of the motivations for our inhomogeneous
cosmological model.

If a smooth power spectrum is assumed for the density perturbations,
it is to be noticed here that the underdense nonlinear structure itself as
a local void on scales 
$\sim 200$ Mpc is a unique object which evolved gravitationally from
primordial adiabatic perturbations but whose amplitude deviates remarkably
from that of average linear perturbations, so that it can be found at
most only a few times within our horizon, and we cannot see anything similar 
 in the neighborhood outside our local void. If similar structures
were found more frequently, we would have a power spectrum with a sharp
spike corresponding to their size.

Owing to the gap of Hubble constants, an inner comoving 
observer has the additional velocity $v_o \ [= 100 (h^{\rm I} - 
h^{\rm II})R_o$ km/s] \ relative to the CMB (outer) frame, if the
distance between the observer and the center of the inner region is 
$R_o$ Mpc. In a previous paper \citep{tom00b} we derived 
the dipole
moment $D$ of CMB anisotropy caused by the additional velocity, and 
the Doppler velocity $v_d$ corresponding to $D$, which depends on 
the model  parameters in the inner and outer regions. If a certain 
condition for the parameters is satisfied, this velocity
$v_d$ gives the above-mentioned difference, so that we may have
\begin{equation}
  \label{eq:r32}
\Vec v_d = 166.6\ {\Vec n}_d \ {\rm km\ s}^{-1},
\end{equation}
where ${\Vec n}_d \equiv (0.925,\ 0.124, \ 0.360)$ \ is also a directional 
unit vector.
 In \S 2 the expressions for the peculiar velocity in the
linear approximation are derived in our model with a local void, and
their difference from the homogeneous case is shown. 
In \S 3 the derivation of $D$ and $v_d$ is briefly accounted for and
their numerical values are shown, and the above condition is discussed.
Some mistakes found in a table in the previous paper are corrected there.
In \S 4, it is shown that the bulk flows can be caused as the sum of
the large-scale peculiar velocity and the additional velocity
$v_o$. Section 5 is dedicated to concluding remarks.
 
\section{Peculiar velocity induced by inhomogeneities}
\label{sec:level2}

Our simple model with a local void consists of the inner homogeneous
region ($V_{\rm I}$) and the outer homogeneous region ($V_{\rm II}$), 
whose boundary is spherical. 
The line elements are expressed as 
\begin{equation}
  \label{eq:m1}
 ds^2 = g^j_{\mu\nu} (dx^j)^\mu (dx^j)^\nu = - c^2 (dt^j)^2 + [a^j
 (t^j)]^2 \Big\{ d (\chi^j)^2 + [f^j (\chi^j)]^2 d\Omega^2 \Big\},
\end{equation}
where $j \ (=$ I or II) represents the regions, $f^j (\chi^j) = \sin
\chi^j, \chi^j$ and $\sinh \chi^j$ for $k^j = 1, 0, -1$, respectively,
and $d\Omega^2 = d\theta^2 + \sin^2 \theta d\varphi^2$.
The boundary shell is a time-like hypersurface $\Sigma$ given as
$\chi^{\rm I} = 
\chi^{\rm I}_1$ and $\chi^{\rm II} = \chi^{\rm II}_1$.

The background models in  $V^{\rm I}$ and $V^{\rm II}$ are specified 
by model parameters $H_0^j, \ \Omega_0^j$ and $\lambda_0^j$ \ for $j = $ 
I and II, where $\Omega_0^j
\equiv [8\pi G / 3(H_0^j)^2] (\rho_0)^j$ \ and \ $\lambda_0^j
\equiv \Lambda
c^2/(H_0^j)^2$, and expressed using $y^j \ (\equiv a^j/(a_0)^j)$ and  
$\tau^j \ (\equiv H_0^j t^j)$ as
\begin{equation}
  \label{eq:m2}
dy^j/d\tau^j = (y^j)^{-1/2} P_j (y^j),  
\end{equation}
where
\begin{equation}
  \label{eq:m3}
\ P_j (y^j)\equiv [\Omega_0^j +
\lambda_0^j (y^j)^3 + (1 - \Omega_0^j - \lambda_0^j) y^j]^{1/2}
\end{equation}
and $a_0^j$ is given by
\begin{equation}
  \label{eq:m4}
a_0^j H_0^j = 1/\sqrt{1 - \Omega_0^j - \lambda_0^j}.
\end{equation}
Their Hubble constants and density parameters in the two regions 
are assumed to satisfy the relations  $H_0^{\rm I} > H_0^{\rm II}$ and
$\Omega_0^{\rm I} < \Omega_0^{\rm II}$. 

In this section we consider local behaviors of matter motion in 
the Newtonian treatment. In terms
of the common time coordinate $t$ and Cartesian coordinates ${\Vec X}$, 
the equations of mass continuity, motion and gravitation are
\begin{equation}
  \label{eq:b1}
{\partial \rho \over \partial t} + \nabla_{\Vec X} (\rho {\Vec V})
= 0,
\end{equation}
\begin{equation}
  \label{eq:b2}
{\partial {\Vec V} \over \partial t} + ({\Vec V}\cdot \nabla_{\Vec X}) 
{\Vec V} + \nabla_{\Vec X}\Phi = 0,
\end{equation}
\begin{equation}
  \label{eq:b3}
{\nabla_{\Vec X}}^2 \ \Phi = 4\pi G \rho,
\end{equation}
where $\rho, {\Vec V}$ and $\Phi$ are the mass density field, the
velocity field and the gravitational potential, and $\rm X$ represents 
the proper distance. The background motion is expressed using uniform
densities $(\rho = \bar{\rho})$ and velocities $({\Vec V} \propto 
{\Vec X})$ in both regions. In order to describe local inhomogeneous motions,
let us next use  for $\Vec X$ the comoving coordinates $\Vec r$
 in the inner region, where ${\Vec X} = a_{\rm I} {\Vec r}$
 and consider linear density and velocity
perturbations $\delta_{\rm I} \ [\equiv (\rho - \bar{\rho}_{\rm
I})/\bar{\rho}_{\rm I}]$ and $\Vec v$ in the inner region, where
$\bar{\rho}_{\rm I}$ is the background (unperturbed) density in the
inner region. Then the first two equations reduce to
\begin{equation}
  \label{eq:b4}
{\partial \delta_{\rm I} \over \partial t} + {1 \over a_{\rm I}}
\nabla_{\Vec r} \cdot {\Vec v} = 0,
\end{equation} 
\begin{equation}
  \label{eq:b5}
{\partial {\Vec v} \over \partial t} + {\dot{a_{\rm I}} \over a_{\rm
I}} {\Vec v} + {1 \over a_{\rm I}}\nabla_{\Vec r} \phi = 0.
\end{equation}
The gravitational equation is
\begin{equation}
  \label{eq:b6}
{1 \over a_{\rm I}^2}{\nabla_{\Vec r}}^2 \phi = 4\pi G (\rho - \bar{\rho}),
\end{equation}
which holds for $\rho$ in both regions and the background density 
$\bar{\rho}$ is represented by $\bar{\rho}_{\rm I}, \bar{\rho}_b$ and
$\bar{\rho}_{\rm II}$ in the 
inner region, the shell, and the outer region, respectively.
The solution for equation (\ref{eq:b6}) is expressed as
\begin{equation}
  \label{eq:b7}
\phi = - G {a_{\rm I}}^2 \int {\bar{\rho}_{\rm I} \delta_{\rm I}  
 \over \vert {\Vec r} - {\Vec r}'\vert} d^3{\Vec r'}
\end{equation}
or
\begin{equation}
  \label{eq:b8}
\phi = - G {a_{\rm I}}^2 \Big[\int_{\rm I} {\bar{\rho}_{\rm I}
\delta_{\rm I} \over \vert {\Vec r} - {\Vec r}'\vert}d^3{\Vec r'} + {1
\over a_{\rm I}} \int_{\rm shell} {(\delta \rho)_b
{r_b}^2 \over \vert {\Vec r} - {\Vec r}_{\rm b}'\vert} \sin \theta' d\theta'
d\varphi' 
+ \int_{\rm II} {(\delta \rho)_{\rm II} \over \vert {\Vec r} - {\Vec
r}'\vert}d^3{\Vec r'} \Big],
\end{equation}
where ${\Vec r}_b$ and $(\delta \rho)_b (\equiv (\rho - \bar{\rho})_b)$
are the coordinate and density perturbation in the boundary shell,
respectively, and
$(\delta \rho)_{\rm II} \equiv \rho - \bar{\rho}_{\rm II}$.

Solving Eqs. (\ref{eq:b4}) and (\ref{eq:b5}), we obtain the velocity
field in the inner region:
\begin{eqnarray}
  \label{eq:b9}
 {\Vec v}({\Vec r}) &=& 
{{a_{\rm I}} \over 4\pi} \int {\dot{\delta}_{\rm I}
({\Vec r} - {\Vec r}') d^3{\Vec r}' \over {\vert {\Vec r} - {\Vec
r}'\vert}^3} + {G \over a_{\rm I}} \int_{\rm shell} \int^t_0 
{a_{\rm I}} (\delta \rho)_b dt 
{{r_b}^2 ({\Vec r} - {\Vec r_b}') \over {\vert {\Vec r} - {\Vec
r_b}'\vert}^3} \sin \theta' d\theta' d\varphi' \cr
&+& {G \over a_{\rm I}} \int_{\rm II} \int^t_0 \bar{\rho}_{\rm I}
{a_{\rm I}}^2 \delta_{\rm I} \Big[{(\delta \rho)_{\rm II} \over 
(\delta \rho)_{\rm I}} - 1\Big] dt
{({\Vec r} - {\Vec r}') \over \vert {\Vec r} - {\Vec r}'\vert^3 } 
d^3{\Vec r}',
\end{eqnarray}
where the first integral in the right-hand side of equation (\ref{eq:b9})  
is performed in both regions and we used the
equation for density perturbation $\delta_{\rm I}$ :
\begin{equation}
  \label{eq:b10}
{\partial^2 \delta_{\rm I}\over \partial t^2} + 2{\dot{a}_{\rm I}\over
a_{\rm I}} {\partial \delta_{\rm I}\over \partial t} = 4\pi
G\bar{\rho}_{\rm I} \delta_{\rm I}.   
\end{equation}
%
%
%
If we adopt the solution $D_{\rm I}$ in the growing
mode in equation (\ref{eq:b10}), and use $f^{\rm I} \equiv d\ln
D_{\rm I}/d\ln a_{\rm I}$, we obtain
\begin{equation}
  \label{eq:b12}
 {\Vec v}({\Vec r}) =
{a_{\rm I} \over 4\pi}H_0^{\rm I}f^{\rm I} \int {{\delta}_{\rm I}
({\Vec r} - {\Vec r}') d^3{\Vec r}' \over {\vert {\Vec r} - 
{\Vec r}'\vert}^3} + \ {Res}, 
\end{equation}
\begin{equation}
  \label{eq:b13}
{Res} \equiv
 \int_{\rm shell} \ J_1 \ {(r_b)^2
({\Vec r} - {\Vec r_b}')  \over \vert {\Vec r} - {\Vec r_b}'\vert^3} 
\sin \theta' d\theta' d\varphi' + 
\int_{\rm II} \ J_2 \ {a_{\rm I} ({\Vec r} - {\Vec r}') \over 
\vert {\Vec r} -{\Vec r}'\vert^3} d^3 {\Vec r}',
\end{equation}
where 
\begin{eqnarray}
  \label{eq:b14}
J_1 &\equiv& {G \over {(a_{\rm I})}} \int^t_0 {a_{\rm I}}
(\delta \rho)_{\rm I} dt,\cr
J_2 &\equiv& {G \over (a_{\rm I})^2}\int^t_0 \Big({(\delta \rho)_{\rm II} 
\over (\delta \rho)_{\rm I}} - 1\Big) \bar{\rho}_{\rm I} {a_{\rm I}}^2 
\delta_{\rm I}dt.
\end{eqnarray}
The first term (A) on the right-hand side of equation (\ref{eq:b12})
represents the peculiar velocity induced in the model that is
homogeneous in the entire region with $H_0^{\rm I}$ and $\Omega_0^{\rm
I}$. The second term ($Res$) gives the contribution of the local
void to the peculiar velocity, which comes from
the inhomogeneous matter distribution. In the homogeneous case without 
Res, the above expression accords with the usual one (e.g. \citet{strauss95}).
The peculiar velocity of the Local Group ${\Vec 
v_{\rm G}}$ is given by the velocity at ${\Vec r} = {\Vec r}_{\rm LG}$ as
${\Vec v}_{\rm G} = {\Vec v}({\Vec r}_{\rm LG})$.

The integrands in A and $Res$  are proportional to 
$ \vert {\Vec r} - {\Vec r}'\vert^{-2}$, and the main contributions in
the integrations A and $Res$ come from the region of distances 
$\sim 40$ Mpc and $> 200$ Mpc, respectively. Accordingly, $Res$ may
be smaller by the factor $5^{-2}$ than A, so that the rough estimate
for the peculiar velocity is given by A and $Res$ is negligible compared
with A. 

Under the assumption that the small-scale density perturbations in the 
luminous matter is $\beta \ \times$ \ those in the dark matter and
$\beta$ is $\approx 1$, \ ${\Vec v}({\Vec r})$ is calculated from 
equation (\ref{eq:b12}) with \ $Res = 0$ by investigating
the inhomogeneity of spatial distributions in galactic redshift surveys.
The values of ${\Vec v}({\Vec r}_{\rm LG})$ were derived by
\citet{rr00} using the PSCz redshift survey including 15,459 galaxies.
When we average their values (in Table 1 of their paper)
over the interval of
$\vert {\Vec r} \vert \ = \ (200$ -- $300) h^{-1}$ Mpc, we obtain
the expression in equation (\ref{eq:b15}).  

\section{Doppler velocity due to the CMB anisotropy for an off-center 
comoving observer}
\label{sec:level3}
In this section we consider the CMB anisotropy that is measured by an
off-center observer in the inner region. For this purpose we must investigate
the behavior of light rays in our model with a local void, by solving 
null-geodesic equations and extracting the multipole components. 
 The detail is described in the previous paper \citep{tom00b}. Here 
we pay attention to light rays that are emitted at the recombination epoch
in the outer region and reached the observer at the present epoch in the 
inner region. The redshift ($z^{\rm II}_{\rm rec}$) corresponding to
the recombination epoch depends on the direction $\phi$ between the light 
rays and $\overrightarrow{\rm CO}$. The value of $z^{\rm II}_{\rm rec}$ is 
numerically  calculated for $0 < \phi < \pi$, and the dipole moment is 
derived.
   
When $z_{\rm rec} \ (= z_{\rm rec}^{\rm II} (\phi))$ is given, the temperature 
$T(\phi)$ of the cosmic background radiation is proportional to $1/(1
+ z_{\rm rec})$, and the dipole moment $D$ and quadrupole moment $Q$ are
defined as 
\begin{eqnarray}
  \label{eq:r28}
D &\equiv& \Bigl\vert\int^\pi_0\int^{2\pi}_0 (1+z_{\rm rec})^{-1}Y_{10} \
\sin \phi d\phi d\varphi\Bigr\vert /\langle (1+z_{\rm rec})^{-1}\rangle,\cr
&=& 2\pi \Bigl\vert\int^\pi_0 (1+z_{\rm rec})^{-1}Y_{10} \
\sin \phi d\phi \Bigr\vert /\langle (1+z_{\rm rec})^{-1}\rangle,
\end{eqnarray}
and 
\begin{eqnarray}
  \label{eq:r29}
Q &\equiv& \Bigl\vert\int^\pi_0\int^{2\pi}_0 (1+z_{\rm rec})^{-1}Y_{20} \
\sin \phi d\phi d\varphi\Bigr\vert /\langle (1+z_{\rm rec})^{-1}\rangle,\cr
&=& 2\pi \Bigl\vert\int^\pi_0 (1+z_{\rm rec})^{-1}Y_{20} \
\sin \phi d\phi \Bigr\vert /\langle (1+z_{\rm rec})^{-1}\rangle,
\end{eqnarray}
where $\langle \rangle$ means the average value taken over the whole sky, and
\begin{eqnarray}
  \label{eq:r30}
Y_{10}(\phi) &=& \sqrt{3 \over 4\pi}\cos \phi, \cr
Y_{20}(\phi) &=& \sqrt{5 \over 4\pi} \Big({3 \over 2}\cos^2 \phi - {1
\over 2}\Big).
\end{eqnarray}
The Doppler velocity $v_d$ corresponding to $D$ is given by
\begin{equation}
  \label{eq:r31}
v_d \equiv c [(3/4\pi)^{1/2} D].
\end{equation}
This velocity $v_d$ is a function of $ R \ (\equiv a_0^{\rm I} \chi_0^{\rm
I}), z_b, (H_0^{\rm I}, \Omega_0^{\rm I})$ and $(H_0^{\rm II},
\Omega_0^{\rm II})$.
Here we fix $H_0^{\rm I}$ as $h^{\rm I} = 0.7$, and fix $z_b$ as 
$z_b = 0.067$, which corresponds to the case 
$r_b (\equiv  a_b^{\rm I} \chi_b^{\rm I}) = 200 (h^{\rm I})^{-1}$
Mpc. In the 6th and 7th columns of Table \ref{table1}, $D$ and $v_d$ 
for $R = 10 
(h^{\rm I})^{-1}$ Mpc are displayed for various model parameters of
the background models. First seven models are spatially flat in the
outer region, and next three models are open. In the inner region all
models are open.  In the last column of Table 1 \ the values of $R \ 
(\equiv R_{170})$ corresponding to $v_d = 170$ km s$^{-1}$ are shown. It is
found from this table that the distance $R$ must be in the interval
$(10$ -- $20)  (h^{\rm I})^{-1}$ Mpc. 

On the other hand, we assume that the Doppler velocity vector 
${\Vec v}_d$ gives the difference
between ${\Vec v}_{\rm G}$ and ${\Vec v}_{\rm CMB}$. From the
comparison between  ${\Vec v}_{\rm G}$ in equation ({\ref{eq:b15}) and ${\Vec
v}_{\rm CMB}$ in equation (\ref{eq:a1}), then the estimated ${\Vec v}_d$ 
is given by equation (\ref{eq:r32}). 
 It is to be noticed that this velocity is in the direction 
of the Galactic center.  

The above derivation of $D$ and $v_d$ is quite the same as that  
in \citet{tom00b}, but it was found that the expressions for $D, Q$ and
$v_d$ in Table 1 of the previous paper have some mistakes with
respect to a factor $2\pi$, and we should read $D (\times 10^4), 
Q (\times 10^4)$ and $v_d$ (km s$^{-1}$) as $D (\times 10^4/2\pi), 
Q (\times 10^4/2\pi)$ and $v_d$ (km s$^{-1}/2\pi$) in Table 1 of
\citet{tom00b}.


\begin{table}
\caption{Dipole moment $D$ and the velocity $v_d$ in the
case $z_b = 0.067$ and $h^{\rm I} = 0.7$.
\label{table1}}
\begin{tabular}{cccccccc}
$\Omega_0^{\rm I}$&$\Omega_0^{\rm II}$&$\lambda_0^{\rm I}$&
$\lambda_0^{\rm II}$&$h^{\rm II}/h^{\rm I}$&$D \ (\times
10^4)$&$v_d$ \ (km/sec)&$R_{170} (\times h^{\rm I})$ Mpc \\
\tableline
0.3& 1.0 & 0.0 & 0.0 & 0.82 & 11.76 &172.8&9.8\tablenotemark{1} \\
0.4& 1.0 & 0.0 & 0.0 & 0.82 & 9.59 &141.0&12.1\tablenotemark{1} \\
0.5& 1.0 & 0.0 & 0.0 & 0.82 & 7.66 &112.6&15.1\tablenotemark{1} \\
0.6& 1.0 & 0.0 & 0.0 & 0.82 & 5.90 &86.7 &19.6\tablenotemark{1} \\
0.3& 0.6 & 0.205 & 0.4 & 0.80 & 7.47 &109.9& 15.5\tablenotemark{1} \\
0.3& 0.6 & 0.303 & 0.4 & 0.87 & 6.72 & 98.9& 17.2\tablenotemark{1} \\
0.4& 0.6 & 0.269 & 0.4 & 0.82 & 5.31 & 78.0& 21.8\tablenotemark{1} \\
0.3& 0.6 & 0.0 & 0.0 & 0.80 & 6.83 &100.4& 16.9\tablenotemark{2} \\
0.3& 0.6 & 0.0 & 0.0 & 0.87 & 6.53 & 95.9& 17.7\tablenotemark{2} \\
0.4& 0.6 & 0.0 & 0.0 & 0.82 & 4.44 & 65.2& 26.1\tablenotemark{2} \\
\end{tabular}
\tablenotetext{1}{The outer space is flat.}
\tablenotetext{2}{The outer space is open.}
\end{table}

\bigskip
\centerline{//\ Table 1\ //}
\bigskip

\section{Large-scale bulk flows}

At each point in the inner region we have a peculiar velocity ${\Vec
v}({\Vec r})$ given by equation (\ref{eq:b12}) that is caused by the
inhomogeneous matter distribution. Averaging the velocity over scales
$\sim r_{\rm bulk}$ longer than $50 h^{-1}$ Mpc, we obtain the bulk flow
velocity ${\Vec v}_b({\Vec r}_{\rm bulk})$. \citet{branch99}
derived the bulk flow velocity for IRAS galaxies in PSCz redshift 
survey including 11206 galaxies, measured out to 
$200 h^{-1}$ Mpc. Its average value is $400 - 500$ km s$^{-1}$ for
$R_{\rm bulk} (\equiv a_0 r_{\rm bulk}) = 60 h^{-1}$ Mpc and 
$350 - 450$ km s$^{-1}$ for $R_{\rm bulk} = 80 h^{-1}$ Mpc. The
amplitude and direction of bulk flows are sensitive to the measured
scales of the matter distribution. If the scales are extended to more
than $300 h^{-1}$ Mpc, the average values of the velocities may increase
over the above ones.

In our inhomogeneous models with a local void, we have not only the
above bulk flow velocity ${\Vec v}_b({\Vec r})$ corresponding to the
peculiar velocity in the region V$^{\rm I}$, but also the velocity
(${\Vec v}_{\rm I}$) of the comoving observer (in V$^{\rm I}$) relative
to the rest frame in V$^{\rm II}$ which is the CMB frame.

The latter velocity ${\Vec v}_{\rm I}$ is equal to $(H_0^{\rm I} 
- H_0^{\rm II}) R_g$ in the direction $\overrightarrow{\rm CG}$ 
from the center to each galaxy G,
where $R_g$ is the distance between C and G. This velocity 
${\Vec v}_{\rm I}$ is expressed as the sum of the radial component 
${\Vec v}_r$ (in the direction $\overrightarrow{\rm OG}$) and the 
constant component 
${\Vec v}_o$ (in the direction $\overrightarrow{\rm CO}$), where $\vert 
{\Vec v}_o\vert =
(H_0^{\rm I} - H_0^{\rm II}) R_o$ and $R_o$ is the distance between C
and O, i.e., ${\Vec v}_{\rm I} = {\Vec v}_r + {\Vec v}_o$. 
The constancy of this component was shown in the previous paper 
\citep{tom00b, tom02}. The constant component ${\Vec v}_o$ behaves as a
peculiar velocity, while the radial component ${\Vec v}_r$ can be
regarded as part of the Hubble motion from the viewpoint of the
observer O. Accordingly, the total observed bulk flow velocity 
${\Vec V}_{\rm bulk}$ is practically expressed using ${\Vec v}_b({\Vec
r}_{\rm bulk})$ and ${\Vec v}_o$ as
\begin{equation}
 \label{eq:f1}
{\Vec V}_{\rm bulk} = {\Vec v}_b({\Vec r}_{\rm bulk}) + {\Vec v}_o. 
\end{equation}
Here let us estimate this velocity, tentatively using one of
the values of Branchini for ${\Vec v}_b({\Vec r}_{\rm bulk})$
on a scale of $60 h^{-1}$ Mpc. Assuming the PSCz M1 method we adopt
\begin{equation}
 \label{eq:f2}
{\Vec v}_b({\Vec r}_{\rm bulk}) = 450\ {\Vec n}_{\rm G} \
{\rm km\ s}^{-1},
\end{equation}
whose direction is approximately taken to be equal to that of 
${\Vec v}_{\rm G}$ derived by \citet{rr00}. We consider the case 
$\Omega_0^{\rm I} = 0.3,  \Omega_0^{\rm II} = 1.0$ and
$\lambda_0^{\rm I} = \lambda_0^{\rm II} = 0$ and $h_0^{\rm II}/
h_0^{\rm I} = 0.82$, which is the best case for representing the SN
data including the data of $z = 1.7$ \ \citep{tom02}. If we take 
$R_o = 10 /h_0^{\rm I}$ \ as in Table 1, we have
\begin{equation}
 \label{eq:f3}
{\Vec v}_o = 180\ \ {\Vec n}_d \ {\rm km\ s}^{-1}.
\end{equation}
Then we obtain from equation (\ref{eq:f1}) 
\begin{equation}
 \label{eq:f4}
{\Vec V}_{\rm bulk} = 378.3\ {\Vec n}_{\rm bulk} \ {\rm km\ s}^{-1},
\end{equation}
where ${\Vec n}_{\rm bulk} \equiv (0.135,\ -0.906,\ 0.402)$ is a 
directional unit vector, so that the total bulk flow velocity has 
the amplitude $378\ {\rm km\ s}^{-1}$, which is $62 \%$ of $v_{\rm CMB}$, 
but its direction is
comparatively close to that of $\Vec v_{\rm CMB}$ in equation (\ref{eq:a1}).

\section{Concluding remarks}
\label{sec:level5}
In this paper we first derived the expression for peculiar velocities
${\Vec v}_b ({\Vec r})$ in the inner region, which are caused
gravitationally by linear matter density perturbations. Next we
derived the Doppler velocity ${\Vec v}_d ({\Vec r})$ corresponding to
the component of the CMB anisotropy caused by a local void. Using
these results we obtained some constraints to our observer's position O 
relative to the center C in the local void, under the assumption that 
${\Vec v}_d$ gives the difference between $v_{\rm CMB}$ and 
${\Vec v}_{\rm G}\ (\equiv {\Vec v} ({\Vec r}_{\rm LG}))$ for the
Local Group.  
It was found that the distance CO must be $(10$ -- $20)/h^{\rm I}$ Mpc, 
depending on the parameters of the background models, and 
$\overrightarrow{\rm CO}$ is in the direction of the Galactic center.

Moreover, we estimated the amplitude and the direction of the bulk
flow velocities on scales of $\sim 60 h^{-1}$ Mpc, as the sum of ${\Vec
v}_b ({\Vec r})$ and the additional flow velocity velocity ${\Vec v}_o$
in the direction of $\overrightarrow{\rm CO}$ which is specified so as 
to give the difference between the inner and outer Hubble constants. 

In our previous treatments \citep{tom00b,tom02} we considered only 
the additional velocity to explain the origin of 
the bulk flow velocities obtained by \citet{hud99}
and \citet{will99}. However, when we consider the peculiar
velocity at the same time, the amplitude of the resultant bulk flow 
velocity is constrained by the observation so as to be much smaller 
than their flow velocities ($\approx 700$ km/s), but the directions
may be roughly consistent.

In \S 5 we neglected $Res$ for the estimation of ${\Vec v}_{\rm G}$, but
because $Res$ is of the order of $4 \%$ of ${\Vec v}_{\rm G}$, it may 
give to ${\Vec v}_{\rm G}$ an error $\sim 25$ km s$^{-1}$, so that it 
may bring an error $\sim 14 \%$ to the difference of the velocities 
$\sim 170$ km s$^{-1}$, or the distance CO ($\sim 10 h^{-1}$ Mpc).

\acknowledgments
This work was supported by Grant-in Aid for Scientific Research 
 12440063 from the Ministry of Education, Science, Sports and
Culture, Japan. 


\newpage
\vspace*{3.cm}
\begin{figure}
\plotone{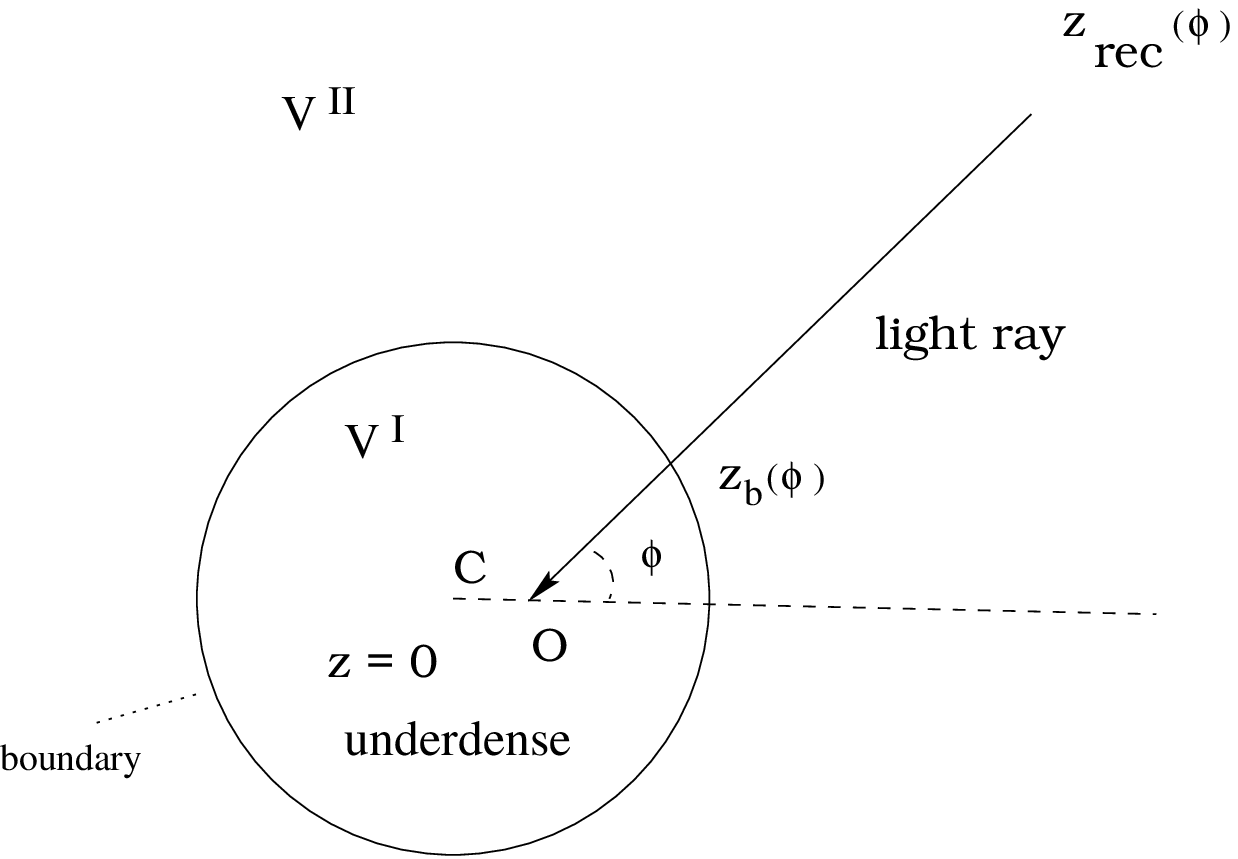}
\caption{A simple model with a spherical boundary. A light ray  is
received by an observer O being near the center C, and $z_b$ and
$z_{\rm rec}$ are the redshifts at the boundary and at the
recombination epoch, respectively, depending on angle $\phi$.  \label{fig1}}
\end{figure}

\end{document}